\documentclass[fleqn,10pt]{wlscirep}
\usepackage[utf8]{inputenc}
\usepackage{array}
\usepackage[T1]{fontenc}
\usepackage{tabu}
\usepackage{mathtools}
\usepackage{bm}

\usepackage{calrsfs}
\DeclareMathAlphabet{\pazocal}{OMS}{zplm}{m}{n}

\usepackage{comment}

\title{Community detection in hypergraphs through hyperedge percolation}

\author[1]{Bianka Kov{\'a}cs}
\author[2]{Barnab{\'a}s Benedek}
\author[2,3,*]{Gergely Palla}
\affil[1]{Institute of Evolution, HUN-REN Centre for Ecological Research, Konkoly-Thege M. {\'u}t 29-33, H-1121 Budapest, Hungary}
\affil[2]{Department of Biological Physics, Eötvös Lor{\'a}nd University, P{\'a}zm{\'a}ny P.\ stny.\ 1/A, H-1117 Budapest, Hungary}
\affil[3]{Health Services Management Training Centre, Semmelweis University, K{\'u}tv{\"o}lgyi {\'u}t 2, H-1125 Budapest, Hungary}

\affil[*]{gergely.palla@emk.semmelweis.hu}

\begin{abstract}

Complex networks often exhibit community structure, with communities corresponding to denser subgraphs in which nodes are closely linked. When modelling systems where interactions extend beyond node pairs to arbitrary numbers of nodes, hypergraphs become necessary, creating a need for specialised community detection methods. Here, we adapt the classical $k$-clique percolation method to hypergraphs, constructing communities from hyperedges containing at least $k$ nodes, defining hyperedge adjacency similarly to clique adjacency. Although the analogy between the proposed hyperedge percolation method and the classical clique percolation algorithm is evident, we show that communities obtained directly from the hyperedges can differ from those identified via clique percolation on the pairwise projection of the hypergraph. We also propose an alternative way for merging hyperedges into communities, where instead of imposing a lower bound on hyperedge cardinality, we restrict the maximum size of the considered hyperedges. This alternative algorithm better suits hypergraphs where larger hyperedges realise weaker linkages between the nodes. After comparing the suggested two approaches on simple synthetic hypergraphs designed to highlight their distinctions, we test them on hypergraphs generated by a newly proposed geometric process on the hyperbolic plane, as well as on some real-world examples.

\end{abstract}
\begin{document}

\flushbottom
\maketitle

\thispagestyle{empty}

\section*{Introduction}
\label{sect:Intro}

Networks are fundamental structures that model complex relationships in various fields, including biology, social sciences, and computer science~\cite{Laci_revmod,Dorog_book,Newman_Barabasi_Watts,Jari_Holme_Phys_Rep,Vespignani_book}. Traditionally, networks have been represented as graphs, where edges represent pairwise (dyadic) interactions. However, in recent years, there has been growing interest in higher-order networks~\cite{Lambiotte_simplicialComplexes,Battison_NetworksBeyondPairwiseInteractions,Eliassi-Rad_higherOrderRepresentations,Bianconi_simplicialComplexesBook,Schaub_higherOrderNetworks,hypergraphx}, such as hypergraphs and simplicial complexes, which extend the classical graph paradigm to capture more complex, group-based (polyadic) interactions among multiple entities. 

A common but non-trivial task to be solved not only for traditional pairwise graphs but also in the case of hypergraphs is the detection of communities (clusters, modules), corresponding to important structural units of a network at the “mesoscopic” scale, usually associated with sub-graphs having a relatively high internal and lower external link density~\cite{Fortunato_coms,Fortunato_Hric_coms,Cherifi_coms}. A vast number of different solutions for revealing the community structure of higher-order networks are already provided in the literature: methods based on statistical inference~\cite{Vazquez_2009,spectralClusteringThenProbabilisticRefinement,maxLikelihoodEstimatorClustering,fromBlockmodelsToModularity,Hypergraph-MT,Hy-MMSBM,mutualInformationMaximizationForClustering,ExpectationMaximizationAlgorithmForClustering}, spectral clustering~\cite{spectralClusteringBasedOnNonbacktrackingHashimotoMatrix,spectralClusteringBasedOnAdjacencyTensor,modularityMaxim2024_spectral}, embedding~\cite{LaplacianEmbeddingAndKmeans,LaplacianEmbeddingAndKmeans2,simplex2vec,tensorEmbedding,nonbacktrackingOperatorEmbeddingAndKmeans}, random walks~\cite{randWalksAndCommDet,flowBasedCommDet} and modularity~\cite{fromBlockmodelsToModularity,modularityMaxim2019,modularityMaxim2020,modularityMaxim2024_faster,modularityMaxim2024_spectral} are all available. 

A successful approach that allows overlaps between the found modules for traditional network systems is provided by the $k$-clique percolation method~\cite{kCliqueComms,CPM_PRL}. Here communities are built from $k$-cliques~\cite{Bollobas_book,Everett_Borgatti_clique_1998}, corresponding to complete sub-graphs consisting of $k$ nodes. Two $k$-cliques are considered to be adjacent if they share $k-1$ nodes, and communities can be defined as maximal sets of $k$-cliques that can be reached from one to the other through a series of adjacent $k$-cliques, making them equivalent to $k$-clique percolation clusters. In the present paper, our main goal is to adapt this idea to hypergraphs, motivated by a natural analogy between cliques and hyperedges, as in some sense both a clique and a hyperedge indicate that everyone is connected to every other member.

Previous work in the literature has already proposed very closely related ideas. Simplicial complexes provide a mathematical framework for the description of higher-order networks that is somewhat more restrictive compared to general hypergraphs. For simplicial complexes, a spectral community finding framework based on the Hodge Laplacian matrix was introduced~\cite{simplicialComms}, where the modules are analogous to $k$-clique communities in traditional networks. Communities based on overlapping hyperedges in general hypergraphs were also introduced~\cite{hyperlinkComms}, where the modules are extracted applying single-linkage hierarchical clustering to the Jaccard distance matrix of the hyperedges. 
Here we aim for communities in general hypergraphs, 
making the community definition fully analogous with the original $k$-clique percolation idea. In addition to the straightforward adaptation of the $k$-clique percolation method to hypergraphs, we also consider an alternative approach, where rather than setting a lower limit on the cardinality of the considered hyperedges, 
we impose an upper limit on the size of hyperedges that can be included in a community.

For validating the proposed two community detection methods, we rely on two real-world hypergraphs and synthetic test frameworks too. On the one hand, we use very simple random modular hypergraphs in which the nodes are classified in pre-defined communities, and the members of inter- and intra-community hyperedges are sampled uniformly at random from all the nodes or from the nodes of a given community, respectively. In addition, we introduce a geometric model for generating hypergraphs. The presence of an underlying hyperbolic geometry has been underpinned by several studies in dyadic graphs~\cite{hyperGeomBasics,Boguna_geometric_weights_ncoms,networkGeometrySummary}, and an emergent hyperbolic geometry has already been connected to models of growing simplicial and cell complexes~\cite{hyperbolicGeometryOfGrowingSimplicalComplexes,hyperbolicGeometryOfGrowingCellComplexes,analyseHyperbolicityOfGrowingSimplicialComplexes}, suggesting that not only pairwise but also higher-order networks might have a natural geometric interpretation. Dyadic graphs generated on the hyperbolic plane using distance-dependent connection probabilities are also exponential random graphs (maximizing ensemble entropy with constraints on the number of links and the sum of their hyperbolic lengths)~\cite{hyperGeomBasics,maximumEntropyApproachOnSimpleGraphs}, 
the family of which has also already gained attention in the context of higher-order networks~\cite{maximumEntropyApproachOnSimplicialComplexes1,Battison_NetworksBeyondPairwiseInteractions,maximumEntropyApproachOnSimplicialComplexes2,maximumEntropyApproachOnHypergraphs}. Here, we propose a hypergraph model that assigns distance-based hyperedges to nodes arranged on the hyperbolic plane and show that our hyperedge percolation methods divide hyperbolic hypergraphs according to angular sectors, in a complete analogy with the observation that angular sectors of the hyperbolic disk correspond to communities in dyadic graphs too~\cite{spatialCommDetOnHypPlane1,spatialCommDetOnHypPlane2,Cannistraci_ASI,inherentCommsOfHypNetworks,Modularity1inRHG,modularity1inPSO}. 

The paper is organised as follows. The Results section first describes the concepts behind the proposed two community detection methods and then demonstrates their operation in random modular hypergraphs, hyperbolic hypergraphs and two real-world datasets. The detailed algorithms for detecting communities based on hyperedge percolation, and for the generation of the simple random modular hypergraphs and the hyperbolic hypergraphs are provided in the Methods section.

\section*{Results}
\label{sect:Results}

\subsection*{Community detection based on hyperedge percolation}

Extending the concept of adjacency from $k$-cliques (i.e., complete subgraphs of $k$ nodes)~\cite{kCliqueComms} to hyperedges, a community in a hypergraph can be defined as a set of nodes being reachable from each other through a series of adjacent hyperedges. In the present study, we propose two community detection approaches based on hyperedge percolation: one that presumes that more localized connections contribute less to the mesoscale structure of the given hypergraph (i.e. the larger a hyperedge, the higher its importance in the community structure), and another assuming that more localized connections correspond to closer, and thus, stronger relationships (i.e., in contrast to the first approach, here the smaller a hyperedge, the more significant role it plays in the communities). The details of the algorithms are given in the Methods section and shown in Fig.~\ref{fig:flowcharts}.

In the first approach, presuming that the mesoscale structure of the given hypergraph is produced by the hyperedges of higher cardinalities, merging the nodes of larger hyperedges in a community is promoted. Accordingly, this method operates on hyperedges containing at least $k$ nodes, bearing a close resemblance to the original $k$-clique percolation method~\cite{kCliqueComms}, an input parameter of which is the minimum size of the $k$-cliques taken into account in the percolation, and thus, in community formation. An advantage of this approach is that this way we can eliminate the excessive inclusion of node sets in cases where small hyperedges are extremely abundant. By following the concept of the original $k$-clique percolation method~\cite{kCliqueComms}, when locating a community, we may imagine a process starting from a hyperedge of size at least $k$, and repeatedly merging it with adjacent hyperedges to form a growing community. In order to get merged into this community, a hyperedge must have at least $I_{\mathrm{abs}}$ number of nodes in common with at least one hyperedge already within the community. 
Since the size of the required absolute intersection is independent of the size of the hyperedge in question, joining a community at a given $I_{\mathrm{abs}}$ is relatively easier for the hyperedges of higher cardinalities. However, increasing the required intersection makes it harder for hyperedges of any cardinalities to join a community. The allowed largest value of $I_{\mathrm{abs}}$ (limiting community growth the most) is $k-1$, which is analogous to the original definition of $k$-clique adjacency~\cite{kCliqueComms}. Nevertheless, according to our investigations, relaxing this definition and testing different values of $I_{\mathrm{abs}}\in[1,k-1]$ can be beneficial. Note that this first approach can be phrased using the concept of higher-order connectivity~\cite{higherOrderComponents} too: the communities detected in a hypergraph correspond to the connected components of order $I_{\mathrm{abs}}$ in a sparsified hypergraph that contains only the hyperedges not smaller than $k$. 

In the second approach, where low-cardinality hyperedges are assumed to be responsible for the formation of communities, the joining of the nodes of smaller hyperedges to a community is favoured over the addition of larger hyperedges. Namely, here we specify an upper limit $K$ on the size of the considered hyperedges, enabling the prevention of the simultaneous inclusion of an excessively large number of nodes to a community based on a single connection, i.e. one extremely large hyperedge. Besides, to enable the inclusion of hyperedges of lower cardinalities even based on a smaller number of nodes in common, and thereby easing the inclusion towards the smaller hyperedge sizes, we condition the community growth on a relative intersection instead of an absolute one. Furthermore, since community growth is more demanding through small hyperedges than through large ones, in order to avoid stopping at too small community sizes, we facilitate community growth in our second algorithm by composing the intersection of a hyperedge in question with the whole of the current community instead of its individual composing hyperedges. All things considered, in our second approach, we expect a hyperedge of size $S$ to have at least $\left \lfloor{I_{\mathrm{rel}}\cdot S}\right \rfloor$ out of its $S$ nodes in common with a given community to get merged in it, where $I_{\mathrm{rel}}\in[0.5,1.0)$ is a tunable parameter of our method.

\subsection*{Finding the planted communities in random modular hypergraphs}

\begin{figure}[h]
    \centering
    \captionsetup{width=\textwidth}
    \includegraphics[width=\textwidth]{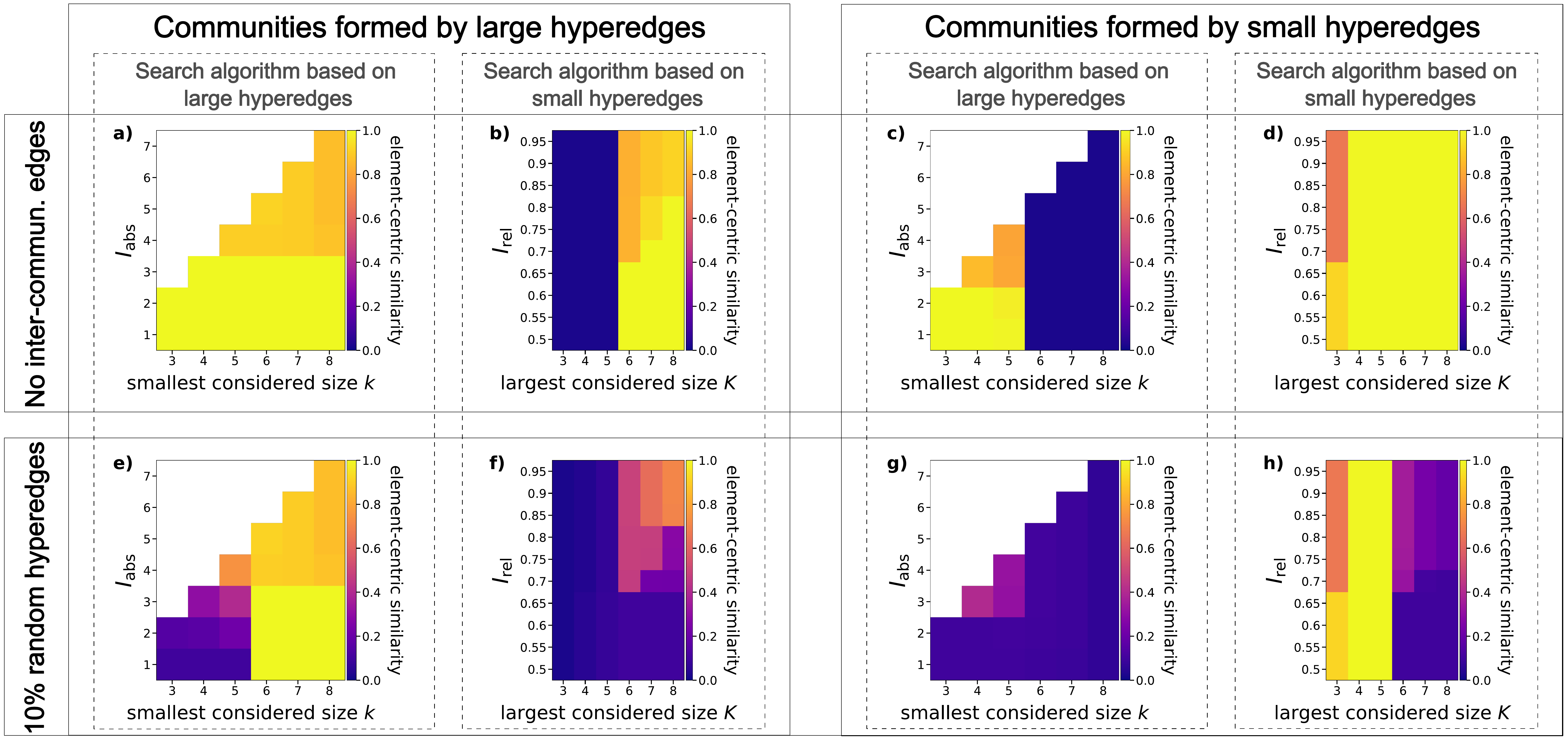}
    \caption{{\bf Community detection performance in random modular hypergraphs.} The four pairs of panels (\textbf{a}--\textbf{b}, \textbf{c}--\textbf{d}, \textbf{e}--\textbf{f} and \textbf{g}--\textbf{h}) deal with four types of hypergraphs consisting of $500$ nodes and $10$ equally sized planted communities. On the left (panels~\textbf{a}, \textbf{b}, \textbf{e} and \textbf{f}), the communities in the examined hypergraphs were formed by large hyperedges (namely, hyperedges of cardinalities $6$, $7$ and $8$, with $50$ of each in each community), while on the right (panels~\textbf{c}, \textbf{d}, \textbf{g} and \textbf{h}), the intra-community hyperedges were small (with cardinalities $3$, $4$ and $5$, $50$ of each in each community). In the hypergraphs investigated in the top row (panels~\textbf{a}--\textbf{d}), there were no inter-community hyperedges, whereas the hypergraphs studied in the bottom row (panels~\textbf{e}--\textbf{h}) contained $50$-$50$-$50$ hyperedges of three cardinality values not used in the community formation (namely $3$, $4$, $5$ in panels \textbf{e}--\textbf{f}, and $6$, $7$, $8$ in panels \textbf{g}--\textbf{h}), connecting nodes that were sampled uniformly at random among all the network nodes. For each of the four examined hypergraph types, a pair of panels depicts the similarity score between the planted community structure and the one identified by the proposed two community finding algorithms, as a function of the community detection parameters: the minimum cardinality of the hyperedges to be considered~($k$) and the absolute intersection of two hyperedges that is required for joining a community~($I_{\mathrm{abs}}$) in the case of the method focusing on large hyperedges (panels~\textbf{a}, \textbf{c}, \textbf{e}, \textbf{g}), or the maximum cardinality of the hyperedges to be considered~($K$) and the relative intersection of a hyperedge and a community that is required for joining~($I_{\mathrm{rel}}$) in the case of the method focusing on small hyperedges (panels~\textbf{b}, \textbf{d}, \textbf{f}, \textbf{h}). All the displayed data correspond to a result averaged over $100$ hypergraph realisations.}
    \label{fig:similarityOnSimpleRandomGraphs}
\end{figure}

To provide some insights into the operation of the proposed community finding algorithms, Fig.~\ref{fig:similarityOnSimpleRandomGraphs} shows their performance as a function of their tunable parameters in the case of some very simple modular hypergraphs. We measured the community detection performance with the element-centric similarity (ECS)~\cite{elCentSim,elCentSimCode} between the planted and the extracted community structures, which is a similarity measure applicable for the comparison of both hard network partitions and overlapping divisions. ECS compares two community structures based on random walks performed along the corresponding two graphs of cluster-induced (i.e., groupmate) relationships, characterising the similarities between the node-node transition probabilities. This measure reaches its maximum of $1$ for identical partitions, and it declines as the similarity between the compared community structures decreases. Note that when inputting two random partitions with an equal number of groups and equal group sizes, the expected value of ECS is not set to $0$~\cite{commDetInNeuralEmbs_Sadamori&Santo}. For non-hierarchical clusterings, the only tunable parameter of ECS is the restart probability of the random walks: in our measurements, we simply used the default value~\cite{elCentSim}, namely $0.10$. 

In the first measurements (Fig.~\ref{fig:similarityOnSimpleRandomGraphs}a--d), we composed hypergraphs from $10$ equally sized, isolated set of nodes, sampling the members of each hyperedge randomly within each group, and then (Fig.~\ref{fig:similarityOnSimpleRandomGraphs}e--h), we added to these hypergraphs hyperedges with members chosen randomly among all the network nodes (see the Methods section for the details of the hypergraph generation). In one test case (Fig.~\ref{fig:similarityOnSimpleRandomGraphs}a,b,e,f), the initial isolated node sets were built up using larger hyperedges that contained at least $6$ nodes and the additional random noise was formed by hyperedges of less than $6$ nodes, whereas in the other test case (Fig.~\ref{fig:similarityOnSimpleRandomGraphs}c,d,g,h) the cardinality of the intra-community hyperedges was below $6$ and the inter-community hyperedges were the larger ones, consisting of at least $6$ nodes. These settings regarding the hyperedge cardinalities ensure that there is a very clear strategy for finding the planted partitions in both cases: any level of random noise interfering the community structure can be completely filtered out by considering only the hyperedges connecting at least $6$ nodes in the first case (Fig.~\ref{fig:similarityOnSimpleRandomGraphs}a,b,e,f), and by disregarding such hyperedges in the second case (Fig.~\ref{fig:similarityOnSimpleRandomGraphs}c,d,g,h). Thus, it is a natural expectation that (if the density of the intra-community hyperedges is large enough for each community) the corresponding variant of hyperedge percolation must be able to exactly recover the planted communities in these hypergraphs.

As it can be seen in Fig.~\ref{fig:similarityOnSimpleRandomGraphs}a--d, in the absence of inter-community hyperedges, both variants of hyperedge percolation reach an ECS of $1$ on both types of hypergraph, meaning perfect partitioning. According to Fig.~\ref{fig:similarityOnSimpleRandomGraphs}a, the algorithm using hyperedges not smaller than $k$ can find the isolated planted communities at any setting of $k$. Since no hyperedges of cardinality $3$, $4$ or $5$ are present, here the examined hyperedge list is the same at $k=3$, $4$ and $5$ as at $k=6$, but above $k=6$, the set of considered hyperedges shrinks, i.e. the considered hypergraph becomes sparser with the increase in $k$. This can narrow the domains available through adjacent hyperedges, making our method divide the planted communities into smaller subsets, as reflected by a decrease in the ECS towards larger values of $k$ at a given $I_{\mathrm{abs}}\geq4$. Meanwhile, in the hypergraphs built up from hyperedges of cardinality $3$, $4$ and $5$ (Fig.~\ref{fig:similarityOnSimpleRandomGraphs}c), $k=6$, $7$ and $8$ means that no hyperedge is taken into account, and thus, no communities are detected, every node assigned to its own community. Here, the maximal ECS of $1$ is achieved only at $k=3$ (i.e., when all the hyperedges of the hypergraph are used) and at $k=4$, while the percolation restricted to the $5$-node hyperedges does not cover the whole communities for any setting of $I_{\mathrm{abs}}$. On this type of hypergraph, our method focusing on hyperedges not larger than $K$ identifies the planted partition with an ECS of $1$ at almost every setting: the only exception is when solely the $3$-node hyperedges are not discarded, i.e. when $K=3$ (see Fig.~\ref{fig:similarityOnSimpleRandomGraphs}d). Note that in the absence of hyperedges containing more than $5$ nodes, the settings $K=6$, $7$ and $8$ are equivalent to $K=5$. However, when the isolated communities are constructed from $6$-, $7$- and $8$-node hyperedges (Fig.~\ref{fig:similarityOnSimpleRandomGraphs}b), the increase in $K$ for $K\geq6$ gradually broadens the set of considered hyperedges, increasing the perceived link density within the planted communities and thereby easing their detection. Without the presence of any $3$-, $4$- and $5$-node hyperedges, the settings $K=3$, $4$ and $5$ mean disregarding all the edges of the hypergraph and classifying each of its nodes in an individual community.

Naturally, the addition of hyperedges that join nodes from any communities makes the detection of the planted communities harder (see Fig.~\ref{fig:similarityOnSimpleRandomGraphs}e--h). Nevertheless, since the random noise is clearly separated from the intra-community hyperedges with regard to the cardinalities in the current simple example, it can be completely eliminated using the proper filtering of the hyperedges. Thus, Fig.~\ref{fig:similarityOnSimpleRandomGraphs}e is the same at $k\geq6$ as Fig.~\ref{fig:similarityOnSimpleRandomGraphs}a, and Fig.~\ref{fig:similarityOnSimpleRandomGraphs}h is the same at $K\leq5$ as Fig.~\ref{fig:similarityOnSimpleRandomGraphs}d. However, when gradually decreasing $k$ below $6$ in Fig.~\ref{fig:similarityOnSimpleRandomGraphs}e or increasing $K$ above $5$ in Fig.~\ref{fig:similarityOnSimpleRandomGraphs}h, as a result of involving more and more noisy connections in the community detection process, the achieved ECS decreases. Obviously, it is also important to choose the detection approach that fits the organising principle of the given community structure in the first place: the noise formed by small hyperedges can never be completely filtered out in the hyperedge percolation method using small hyperedges (Fig.~\ref{fig:similarityOnSimpleRandomGraphs}f), and the noise of large hyperedges always severely burdens the search for communities when focusing on large hyperedges (Fig.~\ref{fig:similarityOnSimpleRandomGraphs}g). Note that measurements at different noise levels and further characteristics of the detected communities are presented in Sect.~S1 of the Supplementary Information.

Decreasing $I_{\mathrm{abs}}$ or $I_{\mathrm{rel}}$ makes the joining of a hyperedge to a growing community easier, while at too high values of these parameters, the hyperedge percolation stops somewhere within the isolated node sets and instead of the planted groups, their subsets are returned as communities. It is important to note that the setting of $I_{\mathrm{abs}}$ that follows the exact concept of clique adjacency proposed in the original clique percolation algorithm is $I_{\mathrm{abs}}=k-1$ (which is also the possible maximum of $I_{\mathrm{abs}}$); nevertheless, as it is also shown by Fig.~\ref{fig:similarityOnSimpleRandomGraphs}, it might be beneficial to not restrict $I_{\mathrm{abs}}$ to a single value and experiment with different settings from the range $[1,k-1]$. Regarding $I_{\mathrm{rel}}$, it can be concluded from Fig.~\ref{fig:similarityOnSimpleRandomGraphs} that its value typically does not affect the community detection performance within relatively large ranges, not necessitating a precise adjustment. 

\subsection*{Community decomposition of hyperbolic hypergraphs}

\begin{figure}[h]
    \centering
    \captionsetup{width=\textwidth}
    \includegraphics[width=\textwidth]{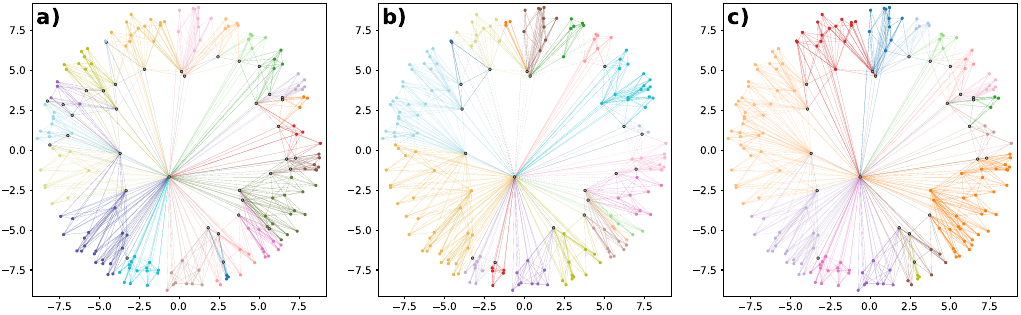}
    \caption{{\bf Communities found in a hyperbolic hypergraph.} The different panels show the partitions detected by different methods on the same hypergraph that was generated on the hyperbolic plane, setting the number of nodes $N$ to $200$, the expected average hyperdegree $\langle k_{\mathrm{h}}\rangle$ to $10$, the $\alpha$ parameter of the radial node distribution to $0.8$, and the hyperedge cardinality distribution to a binomial distribution with an allowed smallest hyperedge cardinality $c_{\mathrm{min}}=3$, an allowed largest hyperedge cardinality $c_{\mathrm{max}}=7$ and an expected average hyperedge cardinality $\langle c\rangle=5$. The depicted node positions correspond to those assigned during the hypergraph generation process. The node colours represent the community memberships. Grey nodes with a black border are assigned to multiple communities (i.e., here different communities overlap), which are indicated by the colour of their links. For the sake of simplicity, instead of hyperedges, the pairwise connections of the hypergraph's clique projection are plotted in each panel. A coloured line or a solid grey line means that the given two connected nodes have exactly one joint community or multiple communities in common, respectively. A dashed grey line represents inter-community links (connecting nodes with no joint community). Panels~\textbf{a} and \textbf{b} present the output of the proposed hyperedge percolation methods when focusing on large and small hyperedges, respectively, whereas panel~\textbf{c} shows the result of the original clique percolation method applied on the hypergraph's dyadic projection. As the size limit of the considered community building blocks, we used the expected average hyperedge cardinality $\langle c\rangle$ in all cases, i.e. the minimum cardinality of the considered hyperedges was $k=5$ in panel~\textbf{a}, the maximum cardinality of the considered hyperedges was $K=5$ in panel~\textbf{b}, and the minimum size of the considered cliques was $k=5$ in panel~\textbf{c}. In panel~\textbf{a}, following the original clique percolation method, we set the required absolute intersection $I_{\mathrm{abs}}$ to $k-1=4$. In panel~\textbf{b}, the required relative intersection $I_{\mathrm{rel}}$ was set to $0.75$.}
    \label{fig:commsInHyperbolicHypergraphs}
\end{figure}

Network models generating graphs on the hyperbolic (i.e., negatively curved) plane~\cite{hyperGeomBasics,PSO} using distance-based connection probabilities have gained considerable attention in recent years due to their ability to capture many essential features of real-world graphs, including their modular structure~\cite{inherentCommsOfHypNetworks,Modularity1inRHG,modularity1inPSO}. Communities 
emerge in hyperbolic graphs due to the inherent density fluctuations of their links through different spatial regions (which occur even in the case of completely uniform spatial distribution of the network nodes) that yield the appearance of angular sectors of more strongly interconnected nodes, and thus, the automatic formation of communities corresponding to angular sectors can be expected in hyperbolic hypergraphs too. In Fig.~\ref{fig:commsInHyperbolicHypergraphs}, we demonstrate through the example of a hyperbolic hypergraph that the output of the proposed hyperedge percolation methods is in line with the expectations, underpinning their capability to identify community-like mesoscale structures in hypergraphs. The hypergraph presented in Fig.~\ref{fig:commsInHyperbolicHypergraphs} was generated on the hyperbolic plane by the algorithm described in the Methods section, marking the boundaries of higher-order connections with hyperbolic circles. It consists of $200$ nodes and $389$ hyperedges, the cardinality of which ranges from $3$ to $7$ with an average of $5.048$. Figure~\ref{fig:commsInHyperbolicHypergraphs}a illustrates a partition found by the hyperedge percolation algorithm using hyperedges of minimum cardinality $k=5$ and Fig.~\ref{fig:commsInHyperbolicHypergraphs}b depicts the partition obtained from hyperedges of maximum cardinality $K=5$. As expected, the two approaches yield different partitions, but the node sets grouped together occupy well-defined angular regions 
on the hyperbolic disk in both cases. 

Figure~\ref{fig:commsInHyperbolicHypergraphs} also shows that although performing the original $k$-clique percolation method~\cite{kCliqueComms} on the clique projection of a hypergraph might seem to be very similar to applying the hyperedge percolation method on the hypergraph itself, the partitions obtained from these two approaches can differ, indicating the importance of handling higher-order data with actual higher-order approaches. Figure~\ref{fig:commsInHyperbolicHypergraphs}c depicts the partition yielded by the original $k$-clique percolation method, defining each community as a node set available in the clique projection of the hypergraph through adjacent $k$-cliques of size $k=5$. When comparing Fig.~\ref{fig:commsInHyperbolicHypergraphs}c to Fig.~\ref{fig:commsInHyperbolicHypergraphs}a, it is easy to see that the lower-order approach (using cliques instead of the actual hyperedges) leads to a different result compared to the analogous higher-order solution: for example, the two orange communities in Fig.~\ref{fig:commsInHyperbolicHypergraphs}c are divided into four and three different communities in Fig.~\ref{fig:commsInHyperbolicHypergraphs}a, despite using on Fig.~\ref{fig:commsInHyperbolicHypergraphs}a those settings that correspond to the parameters of the clique percolation on Fig.~\ref{fig:commsInHyperbolicHypergraphs}c, namely $k=5$ and $I_{\mathrm{abs}}=k-1=4$. The difference is due to the distorting effect of projecting hyperedges to pairwise connections, which allows for the emergence of complete subgraphs even in the absence of a hyperedge. Since cliques corresponding to actual hyperedges and cliques appearing simply due to a higher density of hyperedges are not distinguishable in the dyadic projection of a hypergraph, the clique percolation method operating on the clique projection might merge groups of nodes that form separated communities according to the hyperedges.

\subsection*{Communities detected in real hypergraphs}

\begin{figure}[h!]
    \centering
    \captionsetup{width=\textwidth}
    \includegraphics[width=0.87\textwidth]{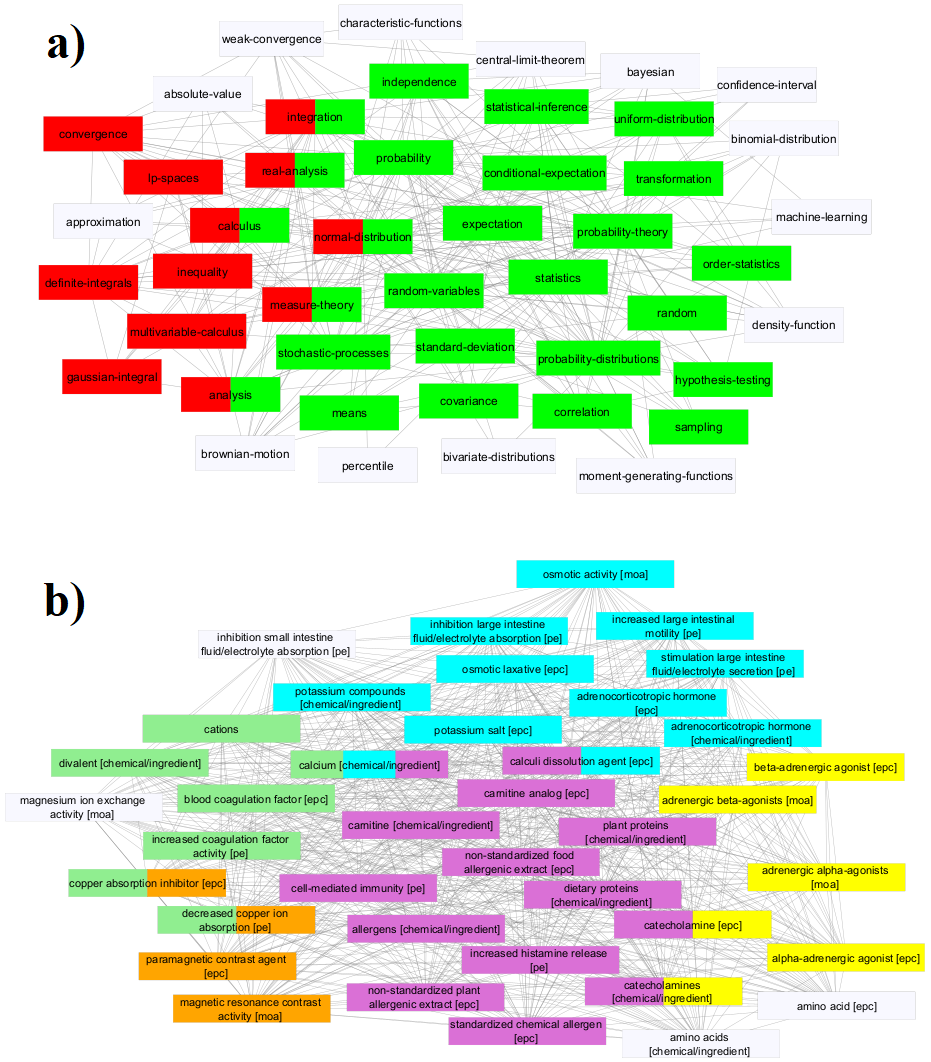}
    \caption{{\bf Communities found in real-world hypergraphs.} Panel~\textbf{a} shows the clique projection of a hypergraph composed of online tagging data, while panel~\textbf{b} depicts the clique projection of a hypergraph obtained from a drug network. The communities detected based on the hyperedges are colour-coded, whereas nodes that were not assigned to any community are displayed in white. The communities of panel~\textbf{a} were identified by the hyperedge percolation method focusing on large hyperedges at $k=5$ and $I_{\mathrm{abs}}=3$, while the partition of panel~\textbf{b} is the output of the hyperedge percolation method using small hyperedges at $K=7$ and $I_{\mathrm{rel}}=0.8$. Independently of the parameter settings, the method focusing on small hyperedges unifies all the nodes of panel~\textbf{a} in a single community, while the method using large hyperedges merges every node in one community when applied on the hypergraph of panel~\textbf{b}.}
    \label{fig:commsInRealHypergraphs}
\end{figure}

As an example of applying hyperedge percolation on real-world data, Fig.~\ref{fig:commsInRealHypergraphs} presents the communities revealed in a hypergraph derived from online tagging data (Fig.~\ref{fig:commsInRealHypergraphs}a) and from a drug network from the National Drug Code (NDC) Directory (Fig.~\ref{fig:commsInRealHypergraphs}b). In the tag hypergraph~\cite{paperWithRealHypergraphs,mathTagData}, the nodes correspond to tags (i.e. annotations or keywords) and a hyperedge contains all the tags that were assigned to a given question on the Mathematics Stack Exchange forum. In the drug hypergraph~\cite{paperWithRealHypergraphs,drugData}, the nodes are the potential class labels of drugs, and a hyperedge connects the set of class labels that are applied to a given drug. 

Although originally both datasets were temporal (including information about when the question was asked and when the drug was first marketed), the timestamps of the hyperedges are simply disregarded in our investigations. Hyperedges occurring multiple times were added only once to our hyperedge lists. In the tag hypergraph, we discarded 
tag combinations that were assigned to only one or two questions and kept only the hyperedges that appeared at least three times in the original dataset. We also omitted the pairwise connections in both datasets and kept only the actual group interactions, i.e. the hyperedges involving at least three nodes. 

Figure~\ref{fig:commsInRealHypergraphs}a shows a subgraph of the tag hypergraph composed from the first neighbours of the node 'normal-distribution'. 
The hyperedges containing nodes outside from the set of the first neighbours were simply truncated, keeping only the members that coappeared with the 'normal-distribution' tag in some questions. 
The result is a hypergraph with $46$ nodes and $704$ hyperedges, among which $443$ have a cardinality of $3$, $210$ include $4$ nodes and $51$ are $5$-node hyperedges. On this hypergraph, the hyperedge percolation method finds two communities when using only the largest hyperedges ($k=5$) and requiring a $3$-node intersection between two hyperedges for joining a community ($I_{\mathrm{abs}}=3$), namely a group of tags of function analysis (coloured red) and probability theory (coloured green), which, naturally, overlap at some nodes. Nodes that do not appear in any of the $5$-node hyperedges are not assigned to either of the two communities (these are displayed in white in Fig.~\ref{fig:commsInRealHypergraphs}a). Due to the very large number of $3$- and also $4$-node hyperedges, the hyperedge percolation focusing on small hyperedges merges all the nodes into a single community at any settings, not being able to reveal any modular structure in this hypergraph. 

Figure~\ref{fig:commsInRealHypergraphs}b deals with the subgraph composed from the nodes encompassed by the largest hyperedge of the drug hypergraph. The hyperedges containing nodes that do not belong to the largest hyperedge were simply truncated, just like in the case of the subgraph of Fig.~\ref{fig:commsInRealHypergraphs}a. 
The result is a hypergraph with $39$ nodes and $67$ hyperedges, the cardinality of which ranges from $3$ to $39$, taking $17$ different values. The absolute frequency of the hyperedges of different sizes lies between $1$ (which is the number of $20$- and $39$-node hyperedges) and $9$ (which is the number of $6$-node hyperedges), but most types of hyperedges occur $2$ times (hyperedges of $3$, $10$, $16$, $17$ and $19$ nodes) or $5$ times (hyperedges of $5$, $8$, $12$, $13$ and $15$ nodes). 
Due to the presence of the $39$-node hyperedge that involves all the nodes of the examined hypergraph, here the hyperedge percolation focusing on large hyperedges returns the whole hypergraph as a single community at any setting. Nevertheless, based on the small hyperedges, the partitioning of this hypergraph becomes possible. Using the $3$-, $4$-, $5$-, $6$- and $7$-node hyperedges ($K=7$), and requiring a relative intersection of $0.8$ between a community and a hyperedge for its joining ($I_{\mathrm{rel}}=0.8$), the hyperedge percolation method divides the hypergraph into $5$ communities, as shown by Fig.~\ref{fig:commsInRealHypergraphs}b. Decreasing $K$ increases the number of nodes that are not assigned to any community, while the growth of $K$ first increases the community overlaps, and then, at $K\geq9$, merges all the nodes in a single community, confirming that in this example the modular structure is grasped by the small hyperedges.

\section*{Discussion}
\label{sect:Discuss}

In this work, we have extended the well-known concept of $k$-clique percolation~\cite{kCliqueComms} to hypergraphs, investigating two very different approaches for community detection, allowing for a fit to communities of various organising principles. Our first method attaches greater importance to larger hyperedges in community formation, while our second algorithm treats rather the smaller hyperedges as the building blocks of communities. Specifying a proper definition for hyperedge adjacency in both cases, we identified communities in hypergraphs as unions of nodes covered by adjacent hyperedges. This process usually yields overlapping communities (classifying some of the nodes to multiple communities at the same time); nevertheless, even traditional hard clusterings can be revealed by our methods if there is a sharp boundary between the community-forming hyperedges and the rest of the connections 
with regard to their cardinality (see Fig.~\ref{fig:similarityOnSimpleRandomGraphs}a,e and Fig.~\ref{fig:similarityOnSimpleRandomGraphs}d,h). 

Though the hyperedge percolation method focusing on large hyperedges can be considered as the direct extension of the original $k$-clique community detection approach~\cite{kCliqueComms}, its output (obtained directly on the hypergraph) frequently differs from the community structure that can be detected on the clique projection of the given hypergraph (see Fig.~\ref{fig:commsInHyperbolicHypergraphs}a,c). In a dyadic projection, many different hyperedges might contribute to the connectivity structure of a given set of nodes and, as converted to cliques, some of the hyperedges might fill in the "gaps" between other, non-adjacent hyperedges. 
Therefore, the original $k$-clique percolation algorithm operating on the pairwise projection of the hypergraph tends to merge together communities that are separated by the analogous hyperedge percolation method. We also note that in the presence of large hyperedges, the projection to pairwise interactions can result in extra large cliques, containing an immense number of pairwise links. This may hinder the location of the traditional $k$-clique percolation clusters in practice, making the presented hypergraph community finding algorithms more advantageous from a technical point of view too. 

In some cases, it might be not obvious which one of the proposed two hyperedge percolation approaches fits a given hypergraph better --- e.g. the partitions shown in both Fig.~\ref{fig:commsInHyperbolicHypergraphs}a and Fig.~\ref{fig:commsInHyperbolicHypergraphs}b seem to be reasonable. Nonetheless, as a rule of thumb, it might be assumed that focusing on high-cardinality hyperedges during community detection is more rewarding in the presence of a disproportionate amount of small hyperedges (as in the case of Fig.~\ref{fig:commsInRealHypergraphs}a), while focusing on low-cardinality hyperedges might be more expedient when there are hyperedges presumably larger than the communities being sought, or even comparable in their size to the total number of nodes in the examined system (as in the case of Fig.~\ref{fig:commsInRealHypergraphs}b).


\section*{Methods}
\label{sect:Methods}

This section first describes the proposed community detection algorithms in detail. Then, it provides the exact description of the processes applied for generating synthetic hypergraphs: the very simplistic approach with the pre-defined community memberships of the nodes and randomly sampled intra- and inter-community hyperedges, and the geometric approach where hyperedges are represented as circles around the network nodes lying on the hyperbolic plane and the communities automatically emerge due to the localised connections in the hypergraph.

\subsection*{Community detection algorithms based on hyperedge percolation}

\begin{figure}[h]
    \centering
    \captionsetup{width=\textwidth}
    \includegraphics[width=\textwidth]{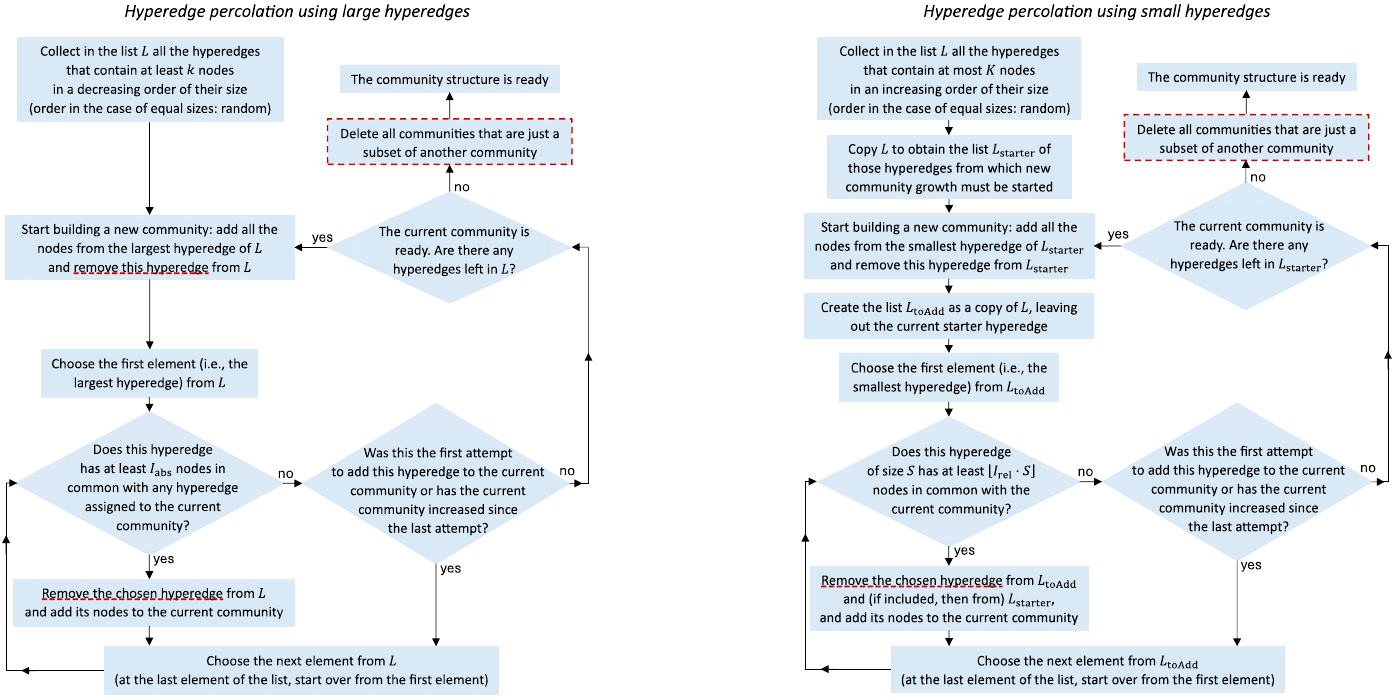}
    \caption{{\bf The proposed community detection algorithms.} The flowchart on the left traces the algorithmic steps of the method focusing on hyperedges of high cardinality, while the flowchart on the right presents the method focusing on hyperedges of low cardinality. The tunable parameters are given on the left by the minimum cardinality of the hyperedges to be considered~($k$) and the absolute intersection of two hyperedges that is required for joining a community~($I_{\mathrm{abs}}\in[1,k-1]$), and on the right by the maximum cardinality of the hyperedges to be considered~($K$) and the relative intersection of a hyperedge and a community that is required for joining~($I_{\mathrm{rel}}\in[0.5,1.0)$). Note that if the list $L$ on the left or $L_{\mathrm{toAdd}}$ on the right becomes empty at any point of the process, then the build-up of the current community is finished at that point and the algorithm jumps to the box framed by red dashed line.}
    \label{fig:flowcharts}
\end{figure}

Figure~\ref{fig:flowcharts} presents the details of the proposed two community detection methods: the left flowchart shows the algorithm for revealing a mesoscale structure that stems from the hyperedges of relatively high cardinalities, while the flowchart on the right specifies the algorithmic steps for partitioning a hypergraph in which the communities rely on the smaller hyperedges. In both cases, a community is identified as a node set that can be covered by a percolation through hyperedges (i.e., by a series of hyperedges that are adjacent to each other in some sense). 
Note that a given set of nodes might be held together by several different sets of hyperedges, and different paths of hyperedge percolation might go through the same node set. Emphasizing the overall transitivity 
between the nodes instead of the different alternatives for going through a set of nodes, we focus on the possible largest clusters that can be covered by a single percolation through the hyperedges. Thus, at the end of both of our algorithms, groups of nodes completely contained by another detected group are discarded. 

To decrease the running time of our algorithms, we reduced the number of times a hyperedge tries to get in a community as much as possible without making the methods non-deterministic. In the algorithm using large hyperedges, where the inclusion rule is symmetric (i.e., if the hyperedge $h_i$ has the required number of nodes in common with the hyperedge $h_j$, then $h_i$ will be merged to any community that contains $h_j$ and vice versa), a hyperedge already assigned to a community does not have to be checked again neither as the seed of a new community nor during the growth of another group. However, in the method focusing on small hyperedges, since the inclusion of a new hyperedge is based on its intersection with the whole set of nodes in the current community instead of individual hyperedges, it is not guaranteed that the joining of a hyperedge $h_i$ to a given community eventually makes the hyperedges similar (or, in some sense, "adjacent") to $h_i$ also joining the community. Therefore, in the algorithm shown on the right of Fig.~\ref{fig:flowcharts}, all the hyperedges must be checked during the growth of every community (thus, the list $L_{\mathrm{toAdd}}$ is initialized in every iteration as a copy of the complete list $L$ of the hyperedges considered in the community detection). Nevertheless, a hyperedge already assigned to a community does not have to be used as a seed of a new community (i.e., the list $L_{\mathrm{starter}}$ can be continuously shortened without losing the deterministic nature of the algorithm), since a new community grown from a hyperedge that is already merged to another community is always a subset of the previous community, and no hyperedge outside of the previous community can be reached in a percolation started from the given hyperedge. 

Note that in the algorithm focusing on large hyperedges, the reasonable settings of the parameter $k$ are larger than the smallest hyperedge cardinality that occurs in the examined hypergraph, and thus, pairwise connections cannot play any role in our first algorithm. However, in our algorithm assuming that the contribution of a hyperedge to the community structure decreases with the hyperedge cardinality, pairwise connections have a special trait: while joining a community can be made harder or easier for any larger hyperedge through the adjustment of $I_{\mathrm{rel}}$, hyperedges of cardinality $2$ face the same requirement at any $I_{\mathrm{rel}}\in[0.5,1.0)$, namely sharing a single node with a given community is already enough for them to join it. Hence, the impact of the pairwise connections on the detected communities cannot be regulated within the algorithm presented on the right of Fig.~\ref{fig:flowcharts}, the inclusion of a $2$-node hyperedge to a community is always very easy here. 
Therefore, when testing our methods on real-world hypergraphs, we discarded the pairwise links and used only the actual higher-order connections, for which the strictness of the inclusion rule is tunable through the size of the required intersection.

\subsection*{A simple generation of random modular hypergraphs}
\label{sect:simplisticHypergraphGeneration}

To construct a synthetic hypergraph with a pre-defined community structure in a very simplistic way, we performed the following steps:
\begin{enumerate}
    \item Classify the $N$ number of nodes to $C$ number of equally sized, non-overlapping communities: assign the community label $i\in[0,C-1]$ to node $j$ for $j\in[i\cdot N/C\,,\,((i+1)\cdot N/C)-1]$.
    \item Build up the binding within the communities: for a given cardinality $S$, create $N/C$ number of hyperedges within each community $i\in[0,C-1]$ by sampling $S$ nodes uniformly at random from $j\in[i\cdot N/C\,,\,((i+1)\cdot N/C)-1]$. 
    \item Generate a random noise of $p$ percentage: for a given cardinality $S$, create $N\cdot p/100$ number of hyperedges by sampling $S$ nodes uniformly at random from all the network nodes, i.e. from $j\in[0,N-1]$.
\end{enumerate}
In our measurements, we always used $3$ different cardinality values for the hyperedges generated both in steps 2 and 3. Note that nothing precludes that some of the hyperedges generated in step 3 will be intra-community hyperedges containing nodes only from a single community, but the majority of the hyperedges of step 3 can be expected to realise inter-community connections, while step 2 generates solely intra-community links.

\subsection*{Hypergraph generation on the hyperbolic plane}
\label{sect:hyperbolicHypergraphGeneration}

Hyperbolic network models, such as the $\mathbb{H}^2$ model~\cite{hyperGeomBasics} and the popularity-similarity optimisation (PSO) model~\cite{PSO}, are well-known for their capability of simultaneously reproducing many universal features of real-world graphs, such as the small-world property
, the relatively high clustering coefficient
, the inhomogeneous degree distribution
, and the presence of communities
. These models place the network nodes on the negatively curved hyperbolic plane and connect them with a distance-dependent probability. In the simplest case (usually referred to as the zero temperature case), the connection rule is deterministic: each node connects to all the nodes within a given hyperbolic distance or to a given number of hyperbolically closest nodes. 

In the native representation~\cite{hyperGeomBasics}, the two-dimensional hyperbolic space of curvature $K<0$ corresponds to a disk of infinite radius in the Euclidean plane, and the node positions are given by usual polar coordinates: $r$ denotes the Euclidean distance from the centre of the disk and $\theta$ stands for the angular coordinate. The hyperbolic distance $x_{ij}$ between two nodes $i$ and $j$ located at polar coordinates $(r_i,\theta_i)$ and $(r_j,\theta_j)$ fulfills the hyperbolic law of cosines:
\begin{equation}
    \mathrm{cosh}(\zeta x_{ij})=\mathrm{cosh}(\zeta r_i)\,\mathrm{cosh}(\zeta r_j)-\mathrm{sinh}(\zeta r_i)\,\mathrm{sinh}(\zeta r_j)\,\mathrm{cos}(\Delta\theta_{ij}),
    \label{eq:hypLawOfCosines}
\end{equation}
where $\zeta=\sqrt{-K}$, and $\Delta\theta_{ij}=\pi-|\pi-|\theta_i-\theta_j||$ is the angular distance between the given two nodes. Note that $r_i=0$ yields $x_{ij}=r_j$, and for $r_j=0$ simply $x_{ij}=r_i$, meaning that in the native representation, the radial coordinate $r$ is equal to not only the Euclidean but also the hyperbolic distance from the disk centre. In the case of $\theta_{ij}=0$, $x_{ij}=|r_i-r_j|$, while for $\theta_{ij}=\pi$, $x_{ij}=r_i+r_j$.

For $2\cdot\sqrt{e^{-2\zeta r_i}+e^{-2\zeta r_j}}<\Delta\theta_{ij}$ and sufficiently large $\zeta r_i$ and $\zeta r_j$, the hyperbolic distance can be approximated as~\cite{hyperGeomBasics}
\begin{equation}
    x_{ij}\approx r_i+r_j+\frac{2}{\zeta}\cdot\ln\left(\frac{\Delta\theta_{ij}}{2}\right).
    \label{eq:hypDistApprox}
\end{equation}
This shows that small hyperbolic distances are yielded by small angular distances and/or small radial coordinates. Thus, the domain that is hyperbolically close to a given point is stretched rather towards the centre of the native disk (radially) than sideways (angularly), toward positions of similarly large radial coordinates. Accordingly, a hyperbolic circle (i.e., the set of points lying at a given hyperbolic distance from a point) is usually drop-shaped --- unless it is centred on the centre of the native disk, in which case the hyperbolic circle looks the same in the native representation as a Euclidean circle. 

\subsubsection*{Hyperbolic model for hypergraph generation}
\label{sect:S1H1_hypergraph}

According to the most common choice, we use the hyperbolic plane of curvature $K=-1$ in our hyperbolic hypergraph model. The tunable parameters of our model can be listed as follows:
\begin{itemize}
    \item The total number of nodes $N$.
    \item The target average hyperdegree $\left< k_{\mathrm{h}}\right>$, corresponding to the expected number of hyperedges containing a randomly chosen node.
    \item The hyperedge cardinality distribution, specifying the expected fraction of hyperedges of different sizes. In our implementation, we consider the following distributions: Dirac delta (parametrised by the single allowed hyperedge size), uniform (parametrised by the allowed smallest and largest hyperedge size), binomial (parametrised by the allowed smallest and largest hyperedge size, and the expected average cardinality), and power-law (parametrised by the allowed smallest and largest hyperedge size, and the decay exponent of the distribution).
    \item The parameter $0.5<\alpha$, controlling the radial distribution of the nodes on the hyperbolic plane. At a given hyperedge size distribution, smaller values of $\alpha$ make larger hubs from the innermost network nodes, yielding more heterogeneous distributions of the node hyperdegrees (corresponding to the number of hyperedges that contain the given node) and the node degrees (corresponding to the total number of nodes to which the given node is connected). Some examples for the obtained degree distributions are shown in Sect.~S2 of the Supplementary Information.
\end{itemize}

From the target average hyperdegree $\left< k_{\mathrm{h}}\right>$ and the expected value $\left< c\right>$ of the inputted hyperedge cardinality distribution, the total number of hyperedges can be calculated as
\begin{equation}
    E = \frac{N\cdot\left< k_{\mathrm{h}}\right>}{\left< c\right>},
    \label{eq:numOfHyperedges}
\end{equation}
rounded to the nearest integer. Then, a hypergraph can be generated through the following two steps:
\begin{enumerate}
    \item All network nodes (indexed by $i=1,2,...,N$) are arranged on the hyperbolic plane the same way as in the $\mathbb{H}^2$ model of pairwise graphs~\cite{hyperGeomBasics}, simply substituting the target average degree in the formulas with the target average hyperdegree $\left< k_{\mathrm{h}}\right>$.
    \begin{enumerate}
        \item For each node $i$, an angular coordinate $\theta_i$ is sampled from $[0,2\pi)$ uniformly at random.
        \item Independently of the angular coordinates, a radial coordinate $r_i$ is sampled for each node $i$ on the native disk~\cite{hyperGeomBasics} from the interval $[0,R]$ according to the probability density function 
        \begin{equation}
        \rho(r)=\alpha\frac{\sinh(\alpha r)}{\cosh(\alpha R)-1},
        \label{eq:radDensity}
        \end{equation}
        where the allowed largest radial coordinate is
        \begin{equation}
        R=2\ln\left(\frac{8N\alpha^2}{\pi\left< k_{\mathrm{h}}\right>(2\alpha-1)^2}\right).
        \label{eq:radOfGraph}
        \end{equation}
    \end{enumerate}
    \item The nodes get connected by $E$ number of different hyperedges. The formation of a hyperedge can be described as follows:
    \begin{enumerate}
        \item Among all network nodes, a node is sampled uniformly at random. This node will be the "centre" of the current hyperedge. 
        \item Independently of the central node, the cardinality $c$ of the current hyperedge is sampled according to the inputted hyperedge cardinality distribution.
        \item A hyperedge connecting the chosen central node and the hyperbolically closest $c-1$ number of nodes is formed. The hyperbolic distance between the central node $i$ and node $j$ is calculated as
        \begin{equation}
        x_{ij}=\mathrm{arccosh}\left(\mathrm{cosh}(r_i)\,\mathrm{cosh}(r_j)-\mathrm{sinh}(r_i)\,\mathrm{sinh}(r_j)\,\mathrm{cos}(\Delta\theta_{ij})\right),
        \label{eq:hypDist}
        \end{equation}
        where $\Delta\theta_{ij}=\pi-|\pi-|\theta_i-\theta_j||$.
        \item If the obtained hyperedge has not appeared previously, then it is added to the edge list of the hypergraph.
    \end{enumerate}
\end{enumerate}
It is important to notice that hyperedges of different central nodes can be the same. In our actual implementation, instead of always re-sampling both the central node and the hyperedge cardinality when the generated hyperedge is already included in the edge list, we create a list of all the necessary $E$ number of cardinality values first and sample the central nodes only afterwards. This way, repeating the generation of hyperedges cannot affect the final hyperedge sizes and even exact lists of hyperedge sizes can be reproduced. Note that the maximum number of different hyperedges of a given size depends on the actual spatial arrangement of the nodes (and is usually smaller than the number $N$ of possible hyperedge centres), and if the required hyperedge cardinality distribution is not feasible in a given hypergraph, then our implementation does not stick strictly to the exact value of $E$ (given by Eq.~(\ref{eq:numOfHyperedges})) but rather tries to adhere to the shape of the required hyperedge size distribution as much as possible, whereas allowing the total number of sampled hyperedge sizes to be larger than $E$ (i.e. the several re-sampling of the hyperedge cardinalities) would yield a better agreement between the obtained average hyperdegree and $\left< k_{\mathrm{h}}\right>$ while increasing the frequency of improbable cardinality values, thereby distorting the hyperedge cardinality distribution. 

Hyperbolic graph models are usually attributed with a tunable temperature parameter, with $T=0$ yielding a deterministic connection rule (where the connection probability is a step function of the hyperbolic distance) and a larger value of $T$ resulting in a less sharp cut-off in the connection probability, larger abundance of long-distance connections, and thus, a decreased clustering coefficient. Our hyperbolic hypergraph model could be also extended 
by not restricting the hyperedge boundaries to regular hyperbolic circles but choosing the members of a hyperedge according to probabilities gradually decreasing with the hyperbolic distance from the "centre" of the given hyperedge (using a Fermi--Dirac distribution~\cite{hyperGeomBasics,PSO}, e.g. setting the cut-off of the connection probability to the hyperbolic distance of the $c-1$th closest neighbour from the given central node). Altering the connection probability function in the hyperbolic hypergraph model is expected to adjust e.g. the nested structure of hyperedges (i.e., how typical the inclusion relationships between hyperedges are), which is an intensively researched property of hypergraphs~\cite{veryBasicInvestigationOfTheNestedStructureOfHyperedges,edgeNestednessOfRealHypergraphs,simplicality,simplicality_newest}. Another possible extension of our model would be the application of nonuniform (multimodal) angular node distributions in order to generate hypergraphs with apparent community structures where planted communities arise at the peaks of the pre-defined angular distribution. For pairwise hyperbolic graphs, models with inhomogeneous angular distributions have already been studied~\cite{GPA_PSOsoftComms,S1softComms,nPSO,nPSO_2}.

\section*{Data availability}
All data generated during the current study are available from the corresponding author upon request. 

\section*{Code availability}
The Python implementation of our two community detection algorithms based on hyperedge percolation, and the Python code used for hypergraph generation on the hyperbolic plane are available at \url{https://github.com/BenedekB17/CommunityDetection-HyperedgePercolation}.

\section*{Acknowledgements}
The research was funded by the National Research, Development and Innovation Office of Hungary (project RRF-2.3.1-21-2022-00006, Data-Driven Health Division of National Laboratory for Health Security).

\section*{Additional information}

\subsection*{Author contributions statement}
B.K. and G.P. developed the concept of the study, B.B. and B.K. implemented the algorithms, B.K. performed the analyses and prepared the figures, B.K. and G.P. wrote the paper. All authors reviewed the manuscript. 


\clearpage

\begin{flushleft} 
{\huge \bfseries{SUPPLEMENTARY INFORMATION}}
\end{flushleft}

\setcounter{section}{0}
\setcounter{equation}{0}
\setcounter{figure}{0}
\setcounter{table}{0}
\renewcommand{\thesection}{S\arabic{section}}
\renewcommand{\thefigure}{S\arabic{figure}}
\renewcommand{\thetable}{S\arabic{table}}
\renewcommand{\theequation}{S\arabic{equation}}

\section{Community detection in random modular hypergraphs}
\label{sect:suppForFig1}

\captionsetup[figure]{font=footnotesize,justification=justified,labelsep=period,labelfont=bf}

\setcounter{figure}{0}
\setcounter{table}{0}
\setcounter{equation}{0}
\renewcommand{\thefigure}{S1.\arabic{figure}}
\renewcommand{\thetable}{S1.\arabic{table}}
\renewcommand{\theequation}{S1.\arabic{equation}}

\begin{figure}[hbt]
    \centering
    \makebox[\textwidth][c]{\includegraphics[width=1.0\textwidth]{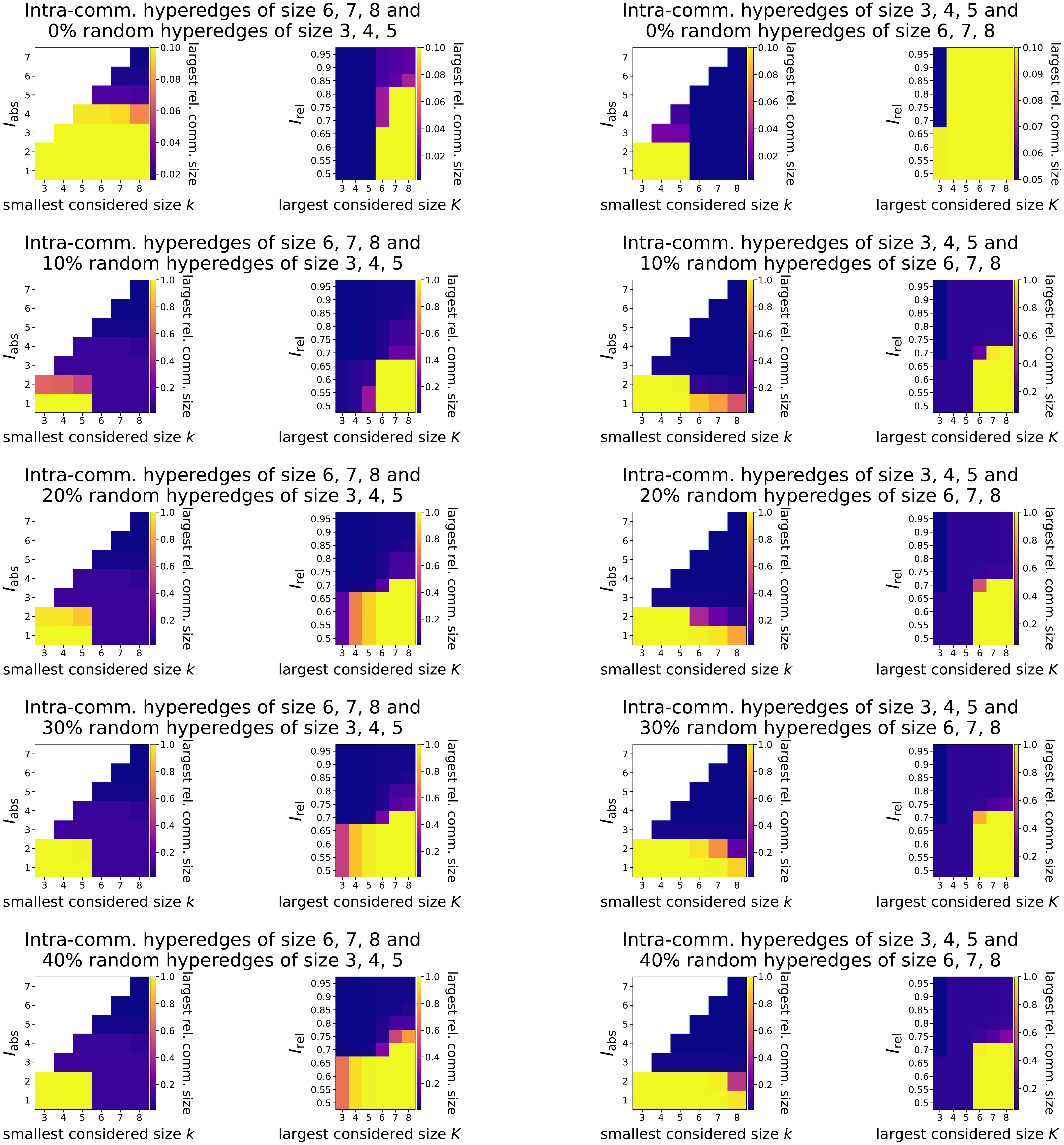}}
    \caption{{\bf The attained largest community size in random modular hypergraphs.} The ten pairs of panels deal with ten types of hypergraphs consisting of $500$ nodes and $10$ equally sized planted communities (i.e., the largest relative community size is $0.1$ in the planted community structure). On the left, the communities in the examined hypergraphs were formed by large hyperedges (of size $6$, $7$ and $8$, with $50$ of each in each community), while on the right, the intra-community hyperedges were small (with cardinalities $3$, $4$ and $5$, $50$ of each in each community). From top to bottom, the ratio of "random noise" (i.e., hyperedges connecting nodes that were sampled uniformly at random among all the network nodes) is increased ($0\%$, $10\%$, $20\%$, $30\%$, $40\%$). The community-forming hyperedges and the completely random hyperedges are always entirely separated in their cardinalities. For each of the ten examined hypergraph types, a pair of panels depicts the relative size of the largest community yielded by the proposed two hyperedge percolation algorithms, as a function of the community detection parameters: the minimum cardinality of the hyperedges to be considered~($k$) and the absolute intersection of two hyperedges that is required for joining a community~($I_{\mathrm{abs}}$) in the algorithm built on large hyperedges, or the maximum cardinality of the hyperedges to be considered~($K$) and the relative intersection of a hyperedge and a community that is required for joining~($I_{\mathrm{rel}}$) in the algorithm built on small hyperedges. All the displayed data correspond to a result averaged over $100$ hypergraph realisations.}
    \label{fig:maxCommSizeInRandModGraphs}
\end{figure}
\begin{figure}[hbt]
    \centering
    \makebox[\textwidth][c]{\includegraphics[width=1.0\textwidth]{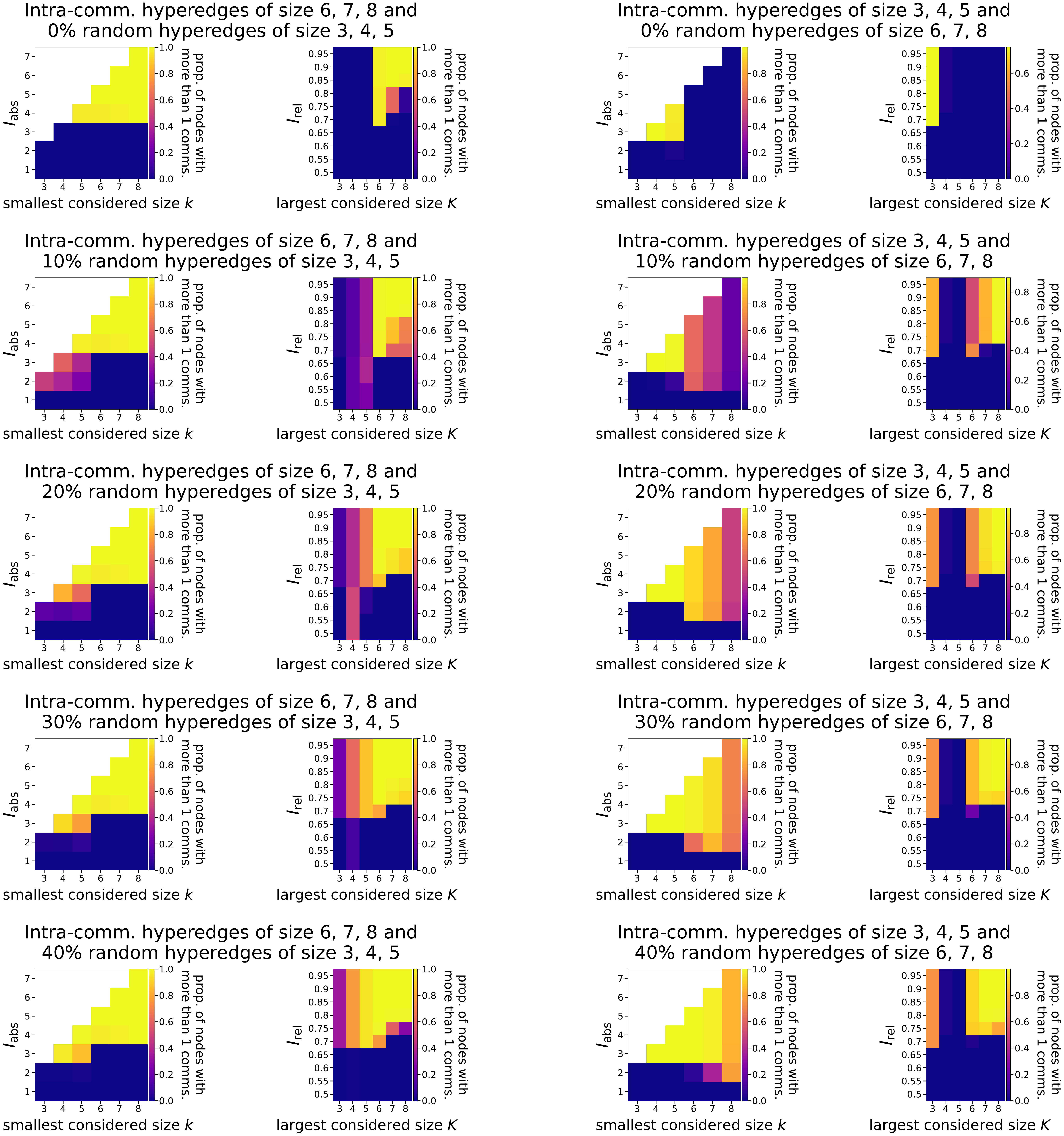}}
    \caption{{\bf Overlaps in the detected community structure of random modular hypergraphs.} The ten pairs of panels deal with ten types of hypergraphs consisting of $500$ nodes and $10$ equally sized planted communities. In the planted community structure, there is no overlap. On the left, the communities in the examined hypergraphs were formed by large hyperedges (of size $6$, $7$ and $8$, with $50$ of each in each community), while on the right, the intra-community hyperedges were small (with cardinalities $3$, $4$ and $5$, $50$ of each in each community). From top to bottom, the ratio of "random noise" (i.e., hyperedges connecting nodes that were sampled uniformly at random among all the network nodes) is increased ($0\%$, $10\%$, $20\%$, $30\%$, $40\%$). The community-forming hyperedges and the completely random hyperedges are always entirely separated in their cardinalities. For each of the ten examined hypergraph types, a pair of panels depicts the proportion of nodes in the network that were classified in more than one community by the proposed two hyperedge percolation algorithms, as a function of the community detection parameters: the minimum cardinality of the hyperedges to be considered~($k$) and the absolute intersection of two hyperedges that is required for joining a community~($I_{\mathrm{abs}}$) in the algorithm built on large hyperedges, or the maximum cardinality of the hyperedges to be considered~($K$) and the relative intersection of a hyperedge and a community that is required for joining~($I_{\mathrm{rel}}$) in the algorithm built on small hyperedges. All the displayed data correspond to a result averaged over $100$ hypergraph realisations.}
    \label{fig:overlapsInRandModGraphs}
\end{figure}
\begin{figure}[hbt]
    \centering
    \makebox[\textwidth][c]{\includegraphics[width=1.0\textwidth]{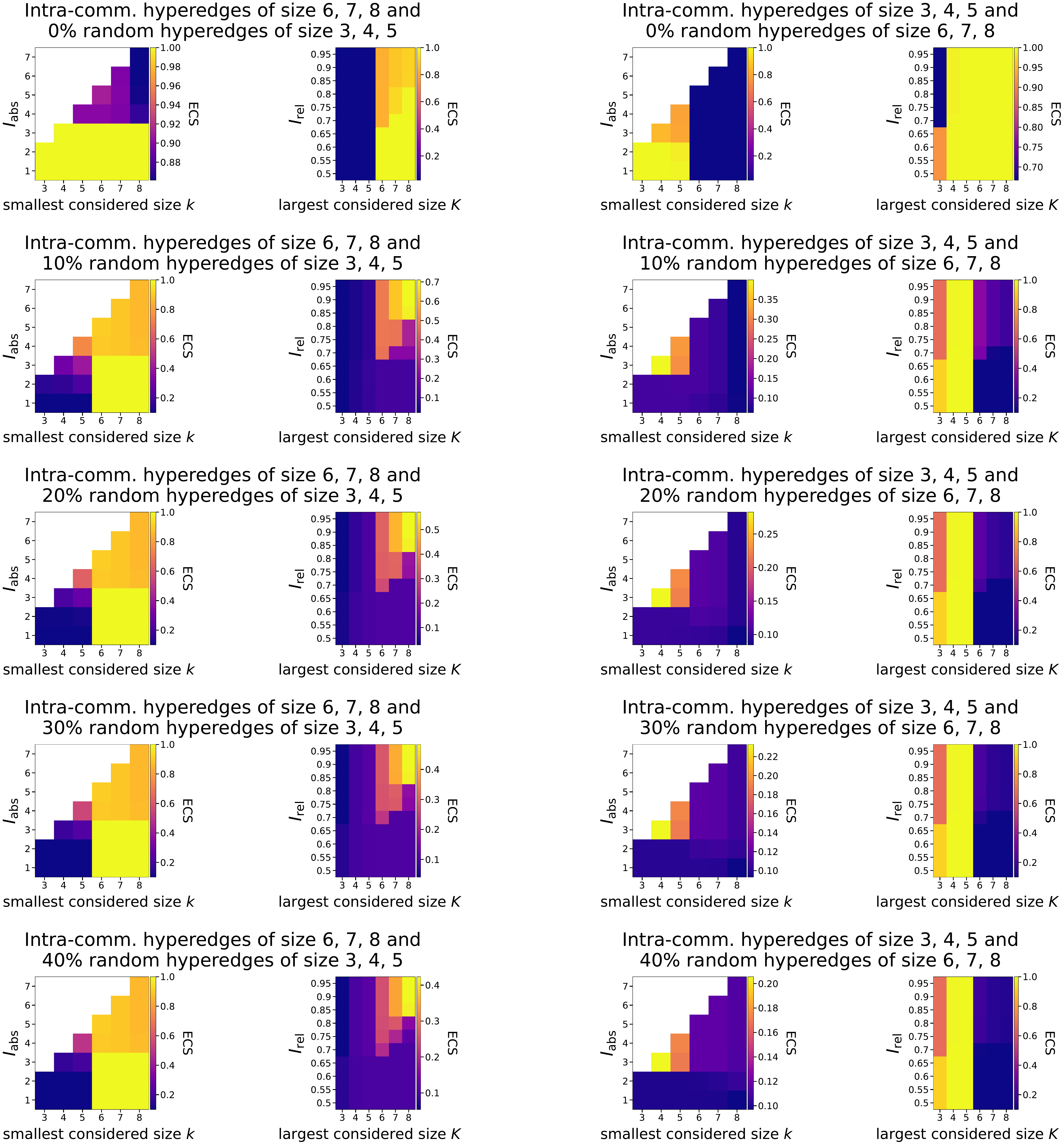}}
    \caption{{\bf Community detection performance in random modular hypergraphs.} The ten pairs of panels deal with ten types of hypergraphs consisting of $500$ nodes and $10$ equally sized planted communities. On the left, the communities in the examined hypergraphs were formed by large hyperedges (of size $6$, $7$ and $8$, with $50$ of each in each community), while on the right, the intra-community hyperedges were small (with cardinalities $3$, $4$ and $5$, $50$ of each in each community). From top to bottom, the ratio of "random noise" (i.e., hyperedges connecting nodes that were sampled uniformly at random among all the network nodes) is increased ($0\%$, $10\%$, $20\%$, $30\%$, $40\%$). The community-forming hyperedges and the completely random hyperedges are always entirely separated in their cardinalities. For each of the ten examined hypergraph types, a pair of panels depicts the element-centric similarity (ECS) between the planted community structure and the one identified by the proposed two hyperedge percolation algorithms, as a function of the community detection parameters: the minimum cardinality of the hyperedges to be considered~($k$) and the absolute intersection of two hyperedges that is required for joining a community~($I_{\mathrm{abs}}$) in the algorithm built on large hyperedges, or the maximum cardinality of the hyperedges to be considered~($K$) and the relative intersection of a hyperedge and a community that is required for joining~($I_{\mathrm{rel}}$) in the algorithm built on small hyperedges. All the displayed data correspond to a result averaged over $100$ hypergraph realisations. The first two rows of panels correspond to Fig.~\ref{fig:similarityOnSimpleRandomGraphs} in the main text.}
    \label{fig:elCentSimInRandModGraphs}
\end{figure}

To elucidate the operation of the proposed community finding algorithms further, this section provides additional insights into their results obtained on simple random modular hypergraphs (described in the Methods~section in the main text and also examined in Fig.~\ref{fig:similarityOnSimpleRandomGraphs} in the main text). Figure~\ref{fig:maxCommSizeInRandModGraphs} presents how limited the community growth is at different parameter settings, displaying the fraction of nodes assigned to the largest detected community. Decreasing the size of the required intersection ($I_{\mathrm{abs}}$ or $I_{\mathrm{rel}}$) makes it easier for hyperedges to join a community, mitigating the limitation of community growth, and thus, increasing the attained largest community size in the detected structures. A higher number of considered hyperedges (due to applying a smaller $k$ or a larger $K$, or simply working with a denser hypergraph obtained at a higher noise level) also enhances the merging of hyperedges and thereby raises the size of the detected largest community. Figure~\ref{fig:overlapsInRandModGraphs} deals with community overlaps, depicting the fraction of network nodes attributed with more than one community label as a function of the tunable parameters of the proposed community finding methods. Here, large fractions indicate that hyperedges barely got merged together, and therefore, most of the detected communities consist of singular hyperedges, easily leading to multiple community memberships for nodes included in more than one hyperedges. Finally, supplementing Fig.~\ref{fig:similarityOnSimpleRandomGraphs} in the main text, Fig.~\ref{fig:elCentSimInRandModGraphs} shows the achieved community detection performances in terms of element-centric similarity (ECS) values~\cite{elCentSim,elCentSimCode} measured between the planted and the extracted community structures.

\section{Degree distribution in hyperbolic hypergraphs}
\label{sect:hyphypDegDist}

\captionsetup[figure]{font=footnotesize,justification=justified,labelsep=period,labelfont=bf}

\setcounter{figure}{0}
\setcounter{table}{0}
\setcounter{equation}{0}
\renewcommand{\thefigure}{S2.\arabic{figure}}
\renewcommand{\thetable}{S2.\arabic{table}}
\renewcommand{\theequation}{S2.\arabic{equation}}

\begin{figure}[hbt]
    \centering
    \makebox[\textwidth][c]{\includegraphics[width=0.985\textwidth]{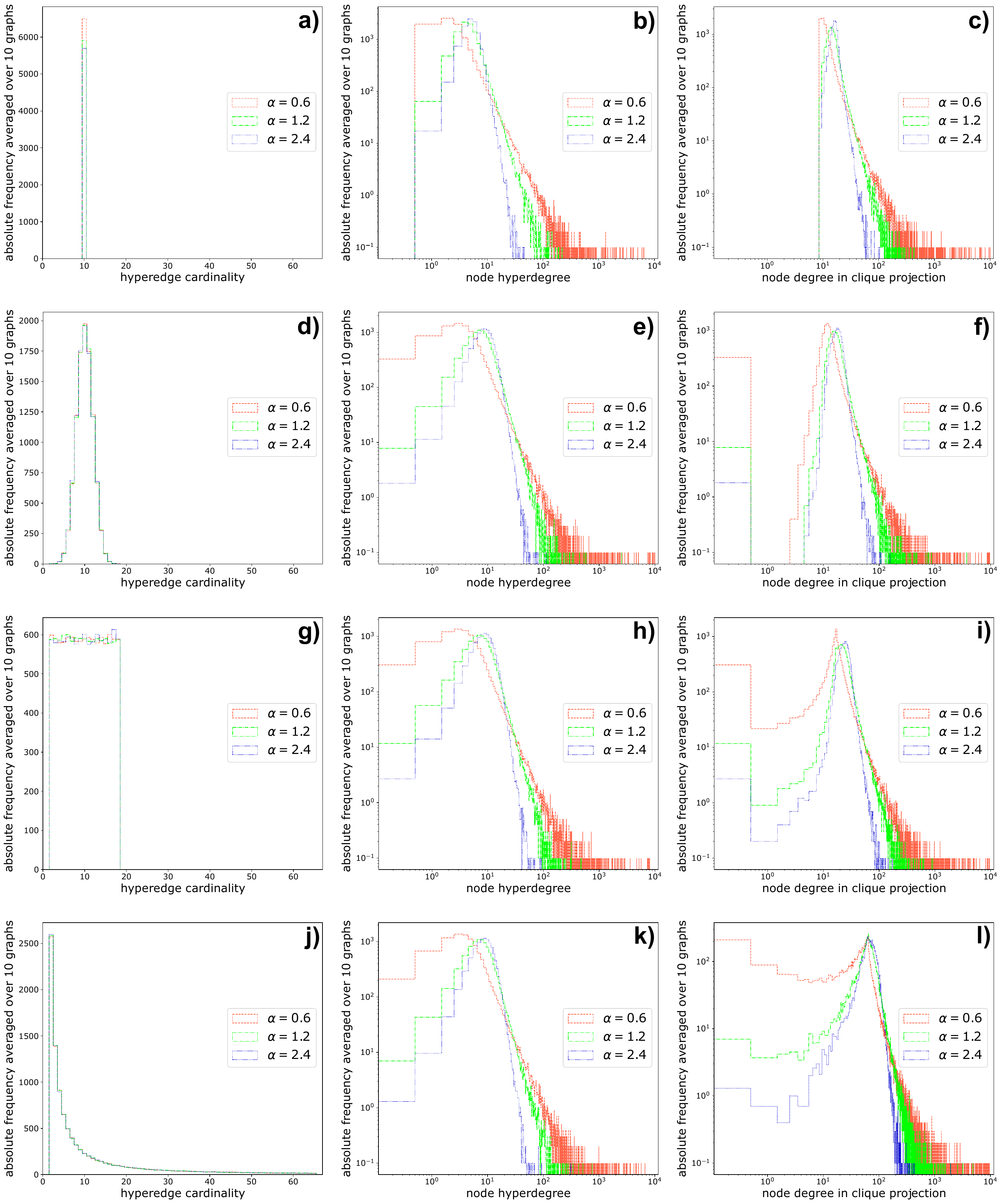}}
    \caption{{\bf The distribution of the size of a hyperedge (left column), the number of hyperedges a node belongs to (middle column), and the total number of nodes a node is connected to (right column) in hypergraphs generated by our hyperbolic hypergraph model.} The examined hypergraphs consist of $N=10000$ nodes, the target average hyperdegree $\left< k_{\mathrm{h}}\right>$ was set to $10$ in all cases, and the depicted data points always correspond to an average over $10$ hypergraphs generated with the same parameter settings. The different colours denote different settings of the $0.5<\alpha$ parameter that controls the radial distribution of the nodes on the hyperbolic plane (a larger value of $\alpha$ increases the empty range in the middle of the native disk and reduces the differences in the radial node coordinates). Each row of panels corresponds to a different setting of the hyperedge cardinality distribution. Panels \textbf{a}--\textbf{c} deal with $10$-uniform hypergraphs, obtained with the Dirac delta edge size distribution, allowing only hyperedges of $10$ nodes. Panels \textbf{d}--\textbf{f} were created with hyperedge cardinalities binomially distributed in the range $[2,18]$. Panels \textbf{g}--\textbf{i} show the case of edge sizes uniformly distributed in the range $[2,18]$. Panels \textbf{j}--\textbf{l} present the results of using power-law decaying hyperedge cardinality distribution, with the decay exponent set to $1.5$ and the range of allowed hyperedge sizes set to $[2,65]$. In panels \textbf{a}--\textbf{i}, the expected value $\left< c\right>$ of the inputted hyperedge cardinality distribution was $10$, while in panels \textbf{j}--\textbf{l}, $\left< c\right>=10.05$.}
    \label{fig:hyphypDegDist}
\end{figure}

As declared in the Methods~section in the main text, the $\alpha$ parameter of the proposed hyperbolic model for hypergraph generation can be used for tuning the heterogeneity of the distributions of the node hyperdegrees and the node degrees by controlling the radial distribution of the network nodes on the hyperbolic plane. Namely, as demonstrated by Fig.~\ref{fig:hyphypDegDist}, lower values of $\alpha$ lead to more heterogeneous (i.e., more slowly decaying) degree distributions at a given setting of the hyperedge cardinality distribution. Note that $\alpha$ usually does not have any impact on the distribution of the hyperedge cardinalities --- the only exception is the case of Dirac delta distribution (Fig.~\ref{fig:hyphypDegDist}a--c), where due to the single allowed hyperedge cardinality, the possible number of different hyperedges is strongly limited (at most $N$, but usually less since not all nodes yield a different hyperedge as a centre) and noticeably affected by the particular node arrangement generated for a given hypergraph, and thus, by the radial node arrangement set by $\alpha$. Nevertheless, except this specific case of Dirac delta hyperedge cardinality distribution, it can be stated that in our hyperbolic hypergraph model, the average hyperdegree $\langle k_{\mathrm{h}} \rangle$ and the hyperedge cardinality distribution are precisely adjustable in a wide range, while at a given setting of these, the decay of the degree distributions can be tuned through $\alpha$. 

\clearpage
\bibliography{references}

\end{document}